# FOURTH- ORDER NONLOCAL TENSOR DECOMPOSITION MODEL FOR SPECTRAL COMPUTED TOMOGRAPHY

*Xiang Chen[1], Wenjun Xia[1], Yan Liu[2], Hu Chen[1], Jiliu Zhou[1], Yi Zhang[1,*]*

[1]College of Computer Science, Sichuan University, Chengdu 610065, China
[2]College of Electrical Engineering, Sichuan University, Chengdu 610065, China

**ABSTRACT**

Spectral computed tomography (CT) can reconstruct spectral images from different energy bins using photon counting detectors (PCDs). However, due to the limited photons and counting rate in the corresponding spectral fraction, the reconstructed spectral images usually suffer from severe noise. In this paper, a fourth-order nonlocal tensor decomposition model for spectral CT image reconstruction (FONT-SIR) method is proposed. Similar patches are collected in both spatial and spectral dimensions simultaneously to form the basic tensor unit. Additionally, principal component analysis (PCA) is applied to extract latent features from the patches for a robust and efficient similarity measure. Then, low-rank and sparsity decomposition is performed on the produced fourth-order tensor unit, and the weighted nuclear norm and total variation (TV) norm are used to enforce the low-rank and sparsity constraints, respectively. The alternating direction method of multipliers (ADMM) is adopted to optimize the objective function. The experimental results with our proposed FONT-SIR demonstrates a superior qualitative and quantitative performance for both simulated and real data sets relative to several state-of-the-art methods, in terms of noise suppression and detail preservation.

## 1. INTRODUCTION

Recently, spectral computed tomography (CT) has attracted considerable interest in several important fields. However, spectral projections acquired from photon counting detectors (PCDs) usually contain strong noise due to limited photons and counting rate in each energy bin. So the development of efficient spectral CT image reconstruction algorithms is of great importance and urgency for clinical applications.

Inspired by the idea of compressed sensing [1], different sparsity constraints, such as total variation [2] and dictionary learning [3], have been proposed for CT image reconstruction and achieved encouraging results. For spectral CT reconstruction, to make use of the correlation among different energy bins, various techniques have been explored to formulate the spectral prior into the reconstruction model. After the work entitled as prior rank, intensity and sparsity model (PRISM), Li *et al.* extended it to a tensor form to further improve the performance [4]. In [3], tensor dictionary learning (TDL), which can be treated as a 3-D extension of classic dictionary learning, was applied to accommodate sparsity and similarity in both spatial and spectral dimensions. Furthermore, in [5], the authors combined TDL and the image gradient $l_0$ norm to achieve better performance in edge preservation. Searching similar image patches in the target image and stacking them into a basic processing unit can efficiently impose the local sparsity and nonlocal similarity simultaneously in a unified framework. Based on this idea, Xia *et al.* extended the model in [6] to a tensor version, whose basic unit is composed of similar patches in both spatial and spectral dimensions [7]. However, since these methods usually vectorize the extracted image patches, the spatial structure and inherent property of high-dimensional data is destroyed and this will lead negative impact on the reconstruction results.

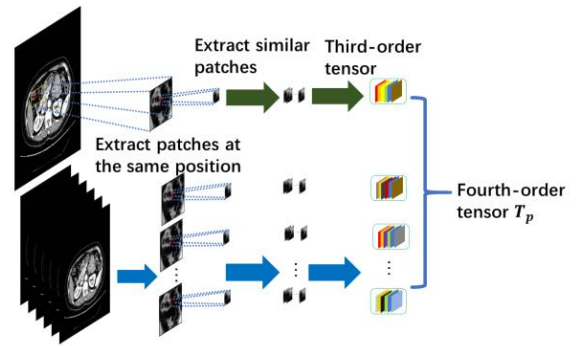

Fig. 1. Illustration of the process to construct a fourth-dimensional tensor unit by spatial and spectral similarities.

To handle the aforementioned issue, in this paper, we propose a Fourth-order Nonlocal Tensor decomposition model for Spectral CT Image Reconstruction (FONT-SIR). First, different from existing tensor-based models, which usually vectorize the 2-D image patches, a basic fourth-order tensor unit is constructed by stacking the similar patches in both spatial and spectral dimensions. In addition, principal component analysis (PCA) is applied to search the similar patches efficiently and robustly [8]. Then low-rank and sparse decomposition is applied to the constructed fourth-order tensor and weighted nuclear norm and total variation norm are respectively adopted to impose the low rank and sparse constraints. Finally, the alternating direction method of multipliers (ADMM) is employed to solve the proposed model.

The rest of the paper is organized as follows. In section 2, the proposed FONT-SIR model and its numerical scheme are presented. Experimental results are given in Section IV and we conclude this paper in the last section.

## 2. METHODOLOGY

In this paper, the proposed spectral CT reconstruction model FONT-SIR can be formulated as

$$min_\chi \frac{1}{2}\|A(\chi) - \zeta\|^2 + \lambda Reg(\chi), \quad (1)$$

where $\chi \in R^{N_h \times N_w \times N_s}$ denotes the spectral CT images as a three-order tensor and $N_s$ is the number of energy bins. $N_h$ and $N_w$ are the image height and width, respectively. $\zeta \in R^{N_d \times N_v \times N_s}$ is the measured data, where $N_d$ and $N_w$ are the numbers of detectors and projection views, respectively. $A(\cdot)$ represents the generalized version of the system matrix for tensors, $Reg(\cdot)$ represents the regularization term, and $\lambda$ is the weighting parameter.

In this work, nonlocal self-similarity in the image is exploited to assist the reconstruction. First, the initial image is obtained by filtered back-projection (FBP). Since the image of the higher energy bin usually has low noise level, the image from the highest energy bin is used as the reference image $X_{ini}$. Second, $N_{num}$ similar patches with size of $N_p \times N_p$ are collected from a $W \times W$ window. Normally, Euclidean distance is employed as the similarity measure. However, it suffers from two main shortcomings: heavy computational costs and inaccurate results once the patches are contaminated by noise and artifacts. To conquer these obstacles, PCA is applied on the patches involved in the comparison. The computation of Euclidean distance directly with the patches is replaced with the projections of patches onto a lower dimensional subspace implemented by PCA. Then, we stack the $N_{num}$ patches into a third-order tensor. Finally, by extracting the third-order tensors in the same locations from all energy bins, $T_p \in R^{N_p \times N_p \times N_{num} \times N_s}$ is obtained. Suppose totally $P$ tensors are collected, the tensor extraction process can be formulated as

$$T_p = GT_p(\chi), p = 1,2, \dots, P \quad (2)$$

where $GT_p(\cdot)$ stands for the tensor extraction operator and $p$ denotes the specific position in the image. The procedure to construct $T_p$ is illustrated in Fig. 1. Once we have all $T_p$, we can decompose $T_p$ into the low-rank component $L_p$ and sparse component $S_p$, respectively. In order to obtain $L_p$ and $S_p$, the minimization problem can be formulated as

$$min_{L_p,S_p} \lambda_1 \|L_p\|_{w,*} + \lambda_2 \|\nabla S_p\|_1, \quad s.t. \ T_p = L_p + S_p \quad (3)$$

TABLE 1: Main Steps of the FONT-SIR Algorithm

| FONT-SIR algorithm |
|---|
| **INPUT:** $A, \zeta, \rho, \lambda_1, \lambda_2, N_{num}, N_p, W$; |
| **Initiation:** |
| Extract patches from the FBP result $\chi_0$ to construct the basic fourth-order tensor unit $T_{p_{p=1}}^P$. |
| **While not meet the stop criterion:** |
|     **For** p=1, 2, …, P |
|         $R_p^n = GT_p(\chi^n)$, |
|         $L_p^{n+1} = argmin_{L_p} \lambda_1 \|L_p\|_{w,*} + \rho/2 \|(T_p^n - S_p^n) - L_p\|^2$, |
|         $S_p^{n+1} = argmin_{S_p} \lambda_2 \|\nabla S_p\|_1 + \rho/2 \|(T_p^n - L_p^{n+1}) - S_p\|^2$, |
|     **End for** |
|     $\chi^{n+1} = argmin_\chi 1/2 \|A(\chi) - \zeta\|^2 + \sum_{p=1}^P \rho/2 \|T_p - L_p^{n+1} - S_p^{n+1}\|^2$, |
|     n=n+1 |
| **End While** |

where $\lambda_1$ and $\lambda_2$ are the balancing parameters. By combining Eq. (1) and (3), the minimization problem can be formulated as

$$min_\chi \frac{1}{2}\|A(\chi) - \zeta\|^2 + \sum_{p=1}^P (\lambda_1 \|L_p\|_{w,*} + \lambda_2 \|\nabla S_p\|_1), \quad (4)$$
$$s.t. \ T_p = GT_p(\chi) = L_p + S_p.$$

The minimization problem in Eq. (4) can be solved using ADMM. The augmented Lagrange form of Eq. (4) can be decomposed into the following sub-problems:

$$L_p^{n+1} = argmin_{L_p} \lambda_1 \|L_p\|_{w,*} + \frac{\rho}{2}\|(T_p^n - S_p^n) - L_p\|^2, \quad (5)$$

$$S_p^{n+1} = argmin_{S_p} \lambda_2 \|\nabla S_p\|_1 + \frac{\rho}{2}\|(T_p^n - L_p^{n+1}) - S_p\|^2, \quad (6)$$

$$\chi^{n+1} = argmin_\chi \frac{1}{2}\|A(\chi) - \zeta\|^2 + \sum_{p=1}^P \frac{\rho}{2}\|T_p - L_p^{n+1} - S_p^{n+1}\|^2. \quad (7)$$

We define $\Omega_k = \text{unfold}_k(T_p^n - S_p^n)$, where $\text{unfold}_k(\cdot)$ represents the operator to unfold the tensor $(T_p^n - S_p^n)$ along the $k$-th mode. Then Eq. (5) can be solved as

$$L_p^{n+1} = \frac{1}{4}\sum_{k=1}^4 \text{fold}_k \left( \sum_i f(\sigma_i^k) u_i^k (v_i^k)^T \right) \quad (8)$$

where $\sigma_i^k$, $u_i^k$ and $v_i^k$ are the singular value and vectors of $\Omega_k$ obtained using singular value decomposition (SVD). $\text{fold}_k(\cdot)$ denotes the inverse operator of $\text{unfold}_k(\cdot)$ to reform the tensor unit. Following the definition of weighted nuclear norm in [9], we have

$$f(\sigma_i^k) = \begin{cases} 0, & if \ c_2 < 0, \\ \frac{c_1 + \sqrt{c_2}}{2}, & if \ c_2 \geq 0, \end{cases} \quad (9)$$

where $c_1 = \sigma_i - \epsilon$ and $c_2 = (\sigma_i + \epsilon)^2 - 4C$. Please refer [9] for more details.

Eq. (6) can be solved by Chambolle projection algorithm [10] as

$$(S_p^{n+1})^m = \xi^m - \mu \ div \ q, \quad (10)$$

where $\xi^m$ and $(S_p^{n+1})^m$ are the $m$-th slice of $(T_p^n - L_p^{n+1})$ and $S_p^{n+1}$ along the 3-rd dimension respectively. $\mu$ is set to $\lambda_2/\rho$ in this paper. $q$ is computed as the following formula

$$q_{i,j,r}^{t+1} = \frac{q_{i,j,r}^t + \tau(\nabla(div\ q^t - \xi^m/\mu))_{i,j,r}}{1 + \tau|\nabla(div\ q^t - \xi^m/\mu)|_{i,j,r}} \quad (11)$$

where $div$ denotes the divergence operator and $\tau$ is the step size. $i, j$, and $r$ represent the indices of the 1-st, 2-nd and 4-th dimension of the tensor, respectively. Once we obtain each $(S_p^{n+1})^m$ with Eq. (10), $S_p^{n+1}$ can be constructed by stacking all the $(S_p^{n+1})^m$.

Once we obtain $L_p^{n+1}$ and $S_p^{n+1}$, the unit $T_p^{n+1} = L_p^{n+1} + S_p^{n+1}$ can be put back to its original position in the image using a general average strategy to form $\tilde{\chi}^{n+1}$ and Eq. (7) is rewritten as

$$min_\chi \frac{1}{2}\|A(\chi) - \zeta\|^2 + \sum_{p=1}^{P}\frac{\rho}{2}\|GT_p(\chi - \tilde{\chi}^{n+1})\|^2. \quad (12)$$

The problem of (12) is convex, which can be solved using conjugate gradient algorithm as

$$A^T(A(\chi)) + \rho \sum_p^P G\ T_p^T\left(GT_p(\chi)\right)$$
$$= A^T(\zeta) + \rho \sum_p^P G\ T_p^T\left(GT_p(\tilde{\chi}^{n+1})\right) \quad (13)$$

where $A^T(\cdot)$ and $GT_p^T(\cdot)$ stand for the transposes of $A(\cdot)$ and $GT_p(\cdot)$, respectively. Since both $A(\cdot)$ and $GT_p(\cdot)$ are linear operators, Eq. (13) is a linear problem, which can be solved by conjugate gradient algorithm. The flowchart of tour method is summarized in Table 1.

## 3. EXPERIMENT AND RESULTS

To validate the performance of the proposed method, both simulated clinical and real data were included in our experiments.

The simulated clinical images were initially obtained using GE Discovery dual-energy CT 750 HD scanner. Virtual monochromic images with size of 256×256 from five different energy bins from 60 to 100keV with every 10 keV increment were produced using GE software. The distance between the scanner rotation center and X-ray source was set to 50 mm. The distance between the radiation detector and X-ray source was set to 100 mm. The detector, whose length was set to 50 mm, was modeled as straight line array with 512 bins. The area of image array covered 20×20 mm². Sparse-view data was used to verify the performance of the proposed FONT-SIR method. 65 views evenly distributed over 360° were chosen. In addition, in order to verify the robustness for low-dose CT reconstruction, following the method in [11], Poisson and electronic noise were added to the simulated projection data.

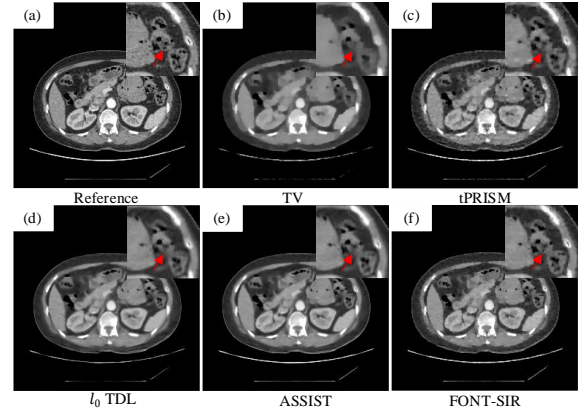

Fig. 2. The results of different methods with simulated noisy abdominal data. There are the results of 80keV energy bin of different methods. The display window is [−160, 240] HU.

The real data was scanned from a mouse injected with gold nanoparticles (GNP). The projection data were acquired by a MARS micro-CT with Medipix MXR CdTe detector. Its X-ray source had a minimum focal spot of 75μm and worked at 120kVp and 175mA. The distance between the X-ray source and the center of rotation was 158 mm. The distance between the X-ray source and the detector was 255 mm. There were totally 512 detector elements with a length of 0.11 mm, and the FOV was 9.21 mm. The X-ray source was divided into 13 energy bins totally. The image size is 256×256, which covers 18.41×18.41 mm². Meanwhile, only 60 views of data from totally 371 views were involved to verify the effectiveness of the method.

In our experiments, we empirically set $\rho = 1 \times 10^5$, , $\lambda_1$=500, $\lambda_2$=40, $W$=11, $N_p$=6, $\tau$=0.125 and $N_{num} = 12$. The stride for patch extraction was set to 3. The values of $\epsilon$ and $C$ were set to $1 \times 10^{-10}$ and $\sqrt{2N_{num}}$. The numbers of iterations for the simulated and real data studies were set to 300 and 30, respectively. Root mean square error (RMSE) and structural similarity index measure (SSIM) were adopted for quantitative assessment. Four state-of-the-art methods, including TV [2], tPRISM [4], $l_0$TDL [5] and ASSIST [7] were included for comparison.

### 3.1. Results on Simulated Data

The results of a noisy abdominal slice using different methods are shown in Fig. 2. The images from 80keV are chosen for comparison. A region of interest (ROI) located in the upper right of the image is magnified for better visualization. The red arrow indicates the location containing features revealed differently by the competing methods. It can be seen that although TV can remove most noise and artifacts, some details are blurred and there are obvious staircase effects due to the mathematical assumption of TV that the signal is piecewise smooth. tPRISM preserves more details than TV, but the artifacts are still noticeable in the result. In the second row of Fig. 2, $l_0$TDL, ASSIST and FONT-SIR suppress most of the noise and artifacts. It can be observed that the proposed FONT-SIR recovers the structural details better than other methods, especially near the bowels.

TABLE 2: RMSE Values in Different Algorithms for Fig. 2

| Methods | TV | tPRISM | $l_0$TDL | ASSIST | FONT-SIR |
|---|---|---|---|---|---|
| 60keV | 0.0087 | 0.0060 | 0.0073 | 0.0051 | **0.0047** |
| 70keV | 0.0078 | 0.0052 | 0.0066 | 0.0041 | **0.0039** |
| 80keV | 0.0075 | 0.0051 | 0.0064 | 0.0041 | **0.0039** |
| 90keV | 0.0073 | 0.0049 | 0.0062 | 0.0039 | **0.0036** |
| 100keV | 0.0069 | 0.0047 | 0.0060 | 0.0037 | **0.0034** |
| Average | 0.0076 | 0.0052 | 0.0065 | 0.0042 | **0.0039** |

TABLE 3: SSIM Values in Different Algorithms for Fig. 2

| Methods | TV | tPRISM | $l_0$TDL | ASSIST | FONT-SIR |
|---|---|---|---|---|---|
| 60keV | 0.9366 | 0.9527 | 0.9740 | 0.9854 | **0.9877** |
| 70keV | 0.9461 | 0.9570 | 0.9790 | 0.9902 | **0.9916** |
| 80keV | 0.9470 | 0.9567 | 0.9800 | 0.9901 | **0.9915** |
| 90keV | 0.9483 | 0.9578 | 0.9815 | 0.9912 | **0.9925** |
| 100keV | 0.9508 | 0.9586 | 0.9828 | 0.9922 | **0.9934** |
| Average | 0.9458 | 0.9566 | 0.9795 | 0.9898 | **0.9913** |

The results from all the energy bins are involved for quantitative comparison. Table 2 and Table 3 respectively show the results of RMSE and SSIM for the reconstructed images in Fig. 2 and other energy bins. We can see that since the lower energy bin usually receives less photons, its result has lower RMSE and SSIM. FONT-SIR achieves the highest scores in terms of both RMSE and SSIM, which is consistent to the visual inspection.

### 3.2. Results on Real Data

We also performed experiments with real preclinical data from a mouse injected with GNP. The images of 7-th channel are chosen for comparison. Fig. 3 shows the reconstruction results with different methods at middle (7-th) energy bins. TV and tPRISM produce remarkable staircase artifacts. $l_0$TDL can remove the noise and artifacts efficiently, but some details are oversmoothed. The upper right parts of the images in Fig. 3 show the magnified ROI indicated by the dotted red box. The red arrow points to a bone structure which can only be well recognized by FONT-SIR. All other methods blur this structure to varying degrees.

### 4. CONCLUSION

In this paper, we propose a fourth-order nonlocal tensor decomposition model for spectral CT image reconstruction. Both simulated and real data were included to evaluate the performance of the proposed FONT-SIR. In the experiments, four different methods, including TV, tPRISM, $l_0$TDL and ASSIST, were compared. In the results, our methods demonstrate better qualitative and quantitative performance than other methods in both noise suppression and detail preservation, which experimentally reveals the effectiveness of our contributions.

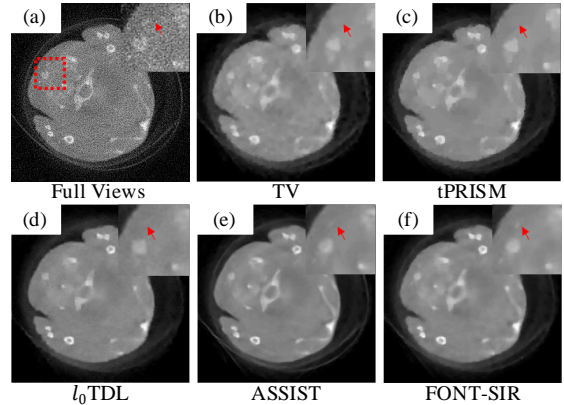

Fig. 3. The results of different methods in true data. There are the results of 7th channel of different methods. The display window of images is [0 0.6] cm$^{-1}$.